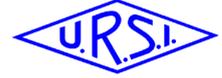

# Third Harmonic Structure in an Interplanetary Type II Radio Burst and Other Energetic Phenomena During the 2024 September 14 Solar Eruption

Nat Gopalswamy*(1), Pertti Makela(2), Hong Xie(2), Sachiko Akiyama, and Seiji Yashiro(2)
(1) NASA Goddard Space Flight Center, Greenbelt, MD, 20771, USA, https://cdaw.gsfc.nasa.gov
(2) The Catholic University of America, Washington DC 20064, USA

## Abstract

We report on the observation of first, second, and third harmonic components during an interplanetary (IP) type II solar radio burst observed on 2024 September 14 by the radio instruments on board Wind, the Solar Terrestrial Relations Observatory (STEREO), and the Parker Solar Probe. The eruption resulted in an ultrafast coronal mass ejection (CME) that had a sky-plane speed of ~2366 km s$^{-1}$ and an X4.5 flare from NOAA active region 13825 (S15E56). Also observed were a large solar energetic particle (SEP) event and a sustained gamma-ray emission (SGRE) event. The IP type II burst consists of multiple features. The first, second, and third harmonic bursts are smooth and diffuse with additional patchy bursts superposed only on the fundamental component. The existence of fundamental-harmonic structure including the third harmonic can be readily explained by the coherent plasma emission mechanism and works against the possibility of synchrotron mechanism.

## 1. Introduction

Observations of plasma emission at harmonics higher than the second harmonic in type II radio bursts from the Sun are extremely rare. Bakunin et al. [1] reported for the first time observations of the first, second, third, and fourth harmonic of a metric type II burst that occurred on 1983 May 12. These authors interpreted that the harmonics originated in closed magnetic traps that had high coronal magnetic field strengths. Kliem et al. [2] presented some evidence for a third harmonic that was observed during the 1984 February 16 metric type II burst (45-330 MHz) associated with a behind-the-limb eruption. These authors considered several mechanisms within the framework of the plasma emission process, including the possibility of strong turbulence. Zlotnik et al. [3] reported three cases of the third harmonic in metric type II bursts. They attributed the rarity of the third harmonic to the non-linear plasma processes responsible for this component. The plasma process involves either the second harmonic radiation combining with a Langmuir wave or three Langmuir waves coalescing to produce the observed third harmonic. Two-dimensional particle-in a-cell (PIC) simulations modeling the propagation of a weak electron beam generating Langmuir waves in solar wind plasmas support the option involving second harmonic radiation combining with a Langmuir wave at the fundamental plasma level [4]. These authors found significant emission at the fourth harmonic as well. Thus, the consensus from observations and simulations is that the harmonic structure is a result of plasma processes resulting from the interaction of energetic electrons with coronal plasma and this happens during energetic eruptions that result in very strong shocks.

In this paper we report on the observations of the first three harmonics in an IP type II radio burst at frequencies below 10 MHz. To our knowledge this is the first time a third harmonic is observed in interplanetary (IP) type II bursts. We also identified a second type II burst (2004 September 12) with the three harmonics and compare it with the 2024 September 14 event.

## 2. Observations

The 2024 September 14 eruption from NOAA active region (AR) AR 13825 located at S15E56 consisted of an intense soft X-ray flare and an ultrafast coronal mass ejection (CME). The X4.5 GOES soft X-ray flare started, peaked and ended at 15:13, 15:29, and 15:47 UT, respectively. The eruption was associated with metric and IP type II bursts, a large solar energetic particle (SEP) event, and a sustained gamma-ray emission (SGRE) event.

The CME was observed by the Large Angle and Spectrometric Coronagraph (LASCO, [5]) on board Solar and Heliospheric observatory (SOHO) and the COR1 and COR2 coronagraphs [6] onboard the Solar Terrestrial Relations Observatory (STEREO). The Radio and Plasma Wave (WAVES) experiment onboard Wind [7] and STEREO [8] observed the type II radio burst. STEREO-A (STA) was located at W25, so the eruption is a limb event in STA view. The SEP event was detected by GOES as a minor (<10 pfu in the >10 MeV channel) but became a large SEP event (≥ 10 pfu) due to the energetic storm particle event around shock arrival at Earth. The shock arrival was observed by Wind, SOHO, STA, and ACE. The shock resulted in a sudden commencement followed by an intense geomagnetic storm (Dst = -121 nT) due to the southward interplanetary magnetic field in the shock sheath (https://cdaw.gsfc.nasa.gov/CME_list/dst100).

The CME appears first in the STA/COR1 field of view (FOV) at 15:21 UT with a leading edge (LE) height of 1.78 Rs and advances to 2.57 RRs and 3.34 Rs at 15:26 UT and 15:31 UT, respectively. In the next frame, the LE already left the FOV. These heights indicate an early speed of ~1786 km s$^{-1}$. Noting that the flare rise time is 16 minutes, initial average acceleration is quite high: 1786/(16×60) =1.9 km s$^{-2}$. Such high initial acceleration and speed are characteristic of energetic CMEs that produce ground level enhancement (GLE) in SEP events and SGRE events [9-11]. The high initial speed can also be confirmed from the first two height-time data points in the LASCO/C2 FOV: 4.81 Rs at 15:36 UT and 5.99 Rs at 15:42 UT, yielding a local speed of 2281 km s$^{-1}$. The CME speed derived from the last two data points: 23.56 Rs at 17:06 UT and 25.76 Rs at 17:18 UT, yielding a local speed of 2126 km s$^{-1}$ (the average speed in the LASCO FOV is 2366 km s$^{-1}$) indicating slight deceleration near the edge of LASCO FOV. All height-time measurements are made available at: https://cdaw.gsfc.nasa.gov/CME_list/UNIVERSAL_ver2/2024_09/yht/20240914.153606.w360h.v2366.p095g.yht.

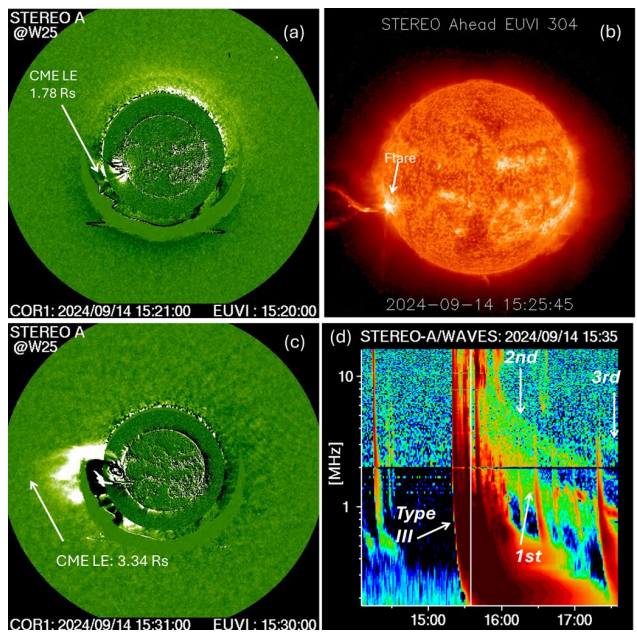

**Figure 1.** (a) The first appearance of the CME in the STA/COR1 FOV at 15:21 UT with the CME leading edge (LE) at 1.78 Rs. (b) An EUV 304 Å image from STEREO showing the flare and eruptive prominence from the east limb in STA view. (c) CME LE advancing to 3.34 Rs at 15:31 UT. (d) STA/WAVES dynamic spectrum showing the type II burst with the three harmonics (1$^{st}$, 2$^{nd}$, 3$^{rd}$). The fundamental (1$^{st}$ harmonic) started at ~5 MHz with second harmonic at ~10 MHz. The third harmonic is very faint. The type II burst lasted from 15:35 UT until ~8:40 UT on the next day (2024 September 15). The bandwidth of the fundamental components is~0.67 measured at 16:00 UT.

Figure 1 shows some snapshots of the eruption imagery. The STA/COR1 image taken at 15:21 UT shows the first appearance of the CME above the east limb at a height of 1.78 Rs (Fig. 1a). The STA/EUVI image at 15:25 UT shows the flare on the east limb and an eruptive prominence above the limb in STA view (Fig. 1b). In the frame at 15:31 UT, the CME LE has advanced to 3.34 Rs (Fig. 1c). A type II burst can be seen in the STA/WAVES dynamic spectrum at 15:35 UT with prominent fundamental-harmonic structure (Fig. 1d). A faint third harmonic is also observed as pointed to by white arrows on the dynamic spectrum. The type II lasts for ~16 hours from the start around 15:30 UT with the three components seen to be present most of this time. The bandwidth of the fundamental measured around 16:00 UT is ~0.67.

Figure 2a shows a vertical cut in the dynamic spectrum at 17:33 UT, close to the edge of the dynamic spectrum in Fig. 1d. The fundamental (1st harmonic) is at 0.85 MHz. The vertical lines in Fig. 2a are at twice (1.7 MHz) and three times (2.6 MHz) this frequency. The three type II peaks agree with the expected frequencies (vertical lines) confirming the three harmonics. The diffuse emissions continued for several more hours ending in the range 6:30 – 8:30 UT on September 15 (see Fig. 2b).

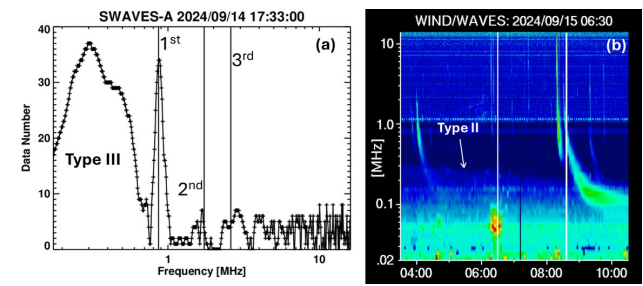

**Figure 2**. (a) A vertical cut at 17:33 UT in the STA/WAVES dynamic spectrum showing the radio signal strength as a function of frequency. The broad peak at low frequencies is the tail of the associated type III burst. The next three peaks correspond to the three harmonics of the type II burst (marked as such). The three vertical lines mark the 1, 2, and 3 times the frequency of the fundamental emission. (b) Wind/WAVES dynamic spectrum showing the end stage of the type II burst. The end time is estimated to be the midpoint (~7:30 UT) between the two vertical lines. The full dynamic spectrum is available in the catalog: https://cdaw.gsfc.nasa.gov/movie/make_javamovie.php?date=20240914&img1=lasc3rdf&img2=wwaves.

The dynamic spectra from ground-based observations show many type III bursts and two possible type II burst episodes in the metric domain. Fig. 3 shows a composite dynamic spectrum that combines e-CALLISTO data from Switzerland with STA/WAVES data. The metric radio emission lasts from ~15:18 to 15:45 UT. Most IP type III bursts have corresponding metric type II bursts. There are two episodes marked "a" and "b" in Figure 3 that are slow-drifting features likely to be type II lanes. The starting frequency of these bursts is relatively low (~60 MHz). It is possible that these type II episodes briefly continue into the SWAVES domain, different from the diffuse component starting at ~10 MHz. Metric type II bursts extending to decameter-hectometric domain are generally thought to originate from the shock flanks [12]. The burst with the triple harmonic is thought to originate from the nose of the shock, which is located at a larger height (and hence lower frequency). Presence of type II emission features in the metric to kilometric wavelengths is a strong indication that the underlying CME is very energetic leading to a strong shock.

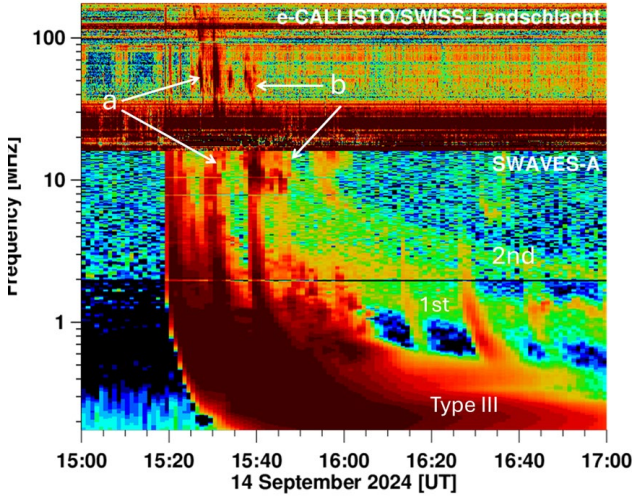

**Figure 3.** A composite dynamic spectrum from ground-based e-CALLISTO network's Landschlacht station in Switzerland (15-180 MHz) and STA/WAVES (0.2-16 MHz). "a" and "b" denote slow-drift features likely to be type II bursts. They also seem to continue into the SWAVES frequency range, The most intense bursts in the SWAVES domain are type III bursts, all of which have metric counterparts.

## 3. Analysis and Results

Based on the rarity of the third harmonic component in metric type II bursts it has been suggested that the underlying shock must be very strong [3]. This is indeed the case during the 2024 September 14 eruption: an ultrafast CME, m-km type II burst, large SEP event, and an SGRE event. Figure 4 shows the timings of the X4.5 flare, CME height-time history, and the intensity of the SEP event an overview of the eruption. Also marked are the durations of the SGRE event and the type II burst. A moderate correlation between soft X-ray flare size and the SEP intensity is known from past studies (e.g., [13]). This correlation can be understood from the fact that energetic CMEs are generally associated with intense soft X-ray flares because both phenomena are connected via the reconnection process that releases stored energy in solar magnetic regions. Therefore, the intense X4.5 flare is associated with an ultrafast CME with a sky-plane speed of 2366 km s$^{-1}$ (deprojected speed of 2854 km s$^{-1}$). As already noted, the type II burst has components from metric to kilometric wavelengths – a property shared by GLE events and SGRE events. The long lifetime indicates that the shock remained strong for large distances from the Sun. The shock arrived at the SOHO spacecraft on September 15 at 16:44 UT indicating a transit time of ~25 hours. This would have been much smaller if the CME were heading directly toward Earth – as small as 14.6 hours, similar to the famous 1972 August 4 event. The SEP event was also long-lasting except that the intensity is <10 pfu in the >10 MeV channel. There were increases in >50 and >100 MeV channels, but the intensities were rather low. The main reason is the poor magnetic connectivity of the eruption to GOES. We can infer that the eruption indeed produced intense high-energy SEP events from the fact that it resulted in an SGRE event, which require the precipitation of >300 MeV protons to the photosphere for almost half a day. These particles, accelerated near the shock nose, do not reach an observer connected to the shock flanks [10]. More details on the SGRE event can be found in the accompanying paper (Gopalswamy et al., AP-RASC 2025).

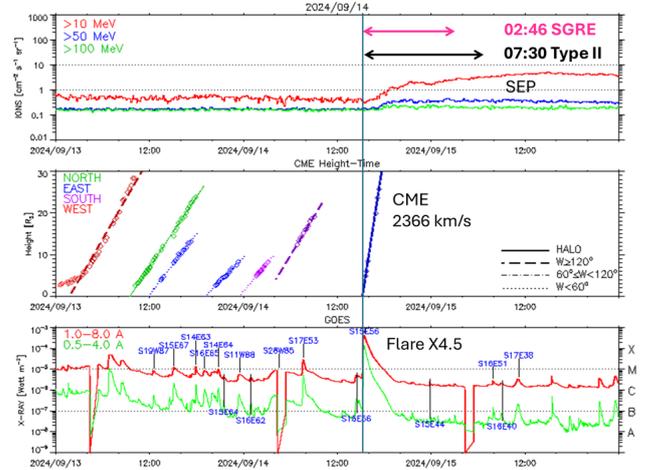

**Figure 4**. (top) GOES proton flux in three integral channels: >10 MeV, >50 MeV, and >100 MeV. (Middle) height-plots of CMEs occurring over a three-day period including the 2024 September 14 CME. (bottom) GOES soft X-ray light curves in the two energy channels including the X4.5 flare in question. The durations of the type II burst and the SGRE event are noted in the top panel.

## 4. Shock Speed from Type II Burst

Under plasma emission hypothesis, one can relate the shock speed (V) with the type II fundamental emission frequency f, the drift rate (df/dt), and the density (n) scale height $L= |n/(dn/dr)|$, $V = 2L(1/f)(df/dt)$. From the type II fundamental component between 16:00 and 18:00 UT in Fig. 5, we get the average drift rate as $2.8\times10^{-4}$ MHz s$^{-1}$. In the frequency range 10 - 1 MHz, the electron density distribution can be approximated by a power law with an index (α) of 4 [12], so the density scale height at a heliocentric distance r becomes L = r/α. The fundamental emission at 1 MHz occurs at 17:00 UT, when the CME LE is >23 Rs. We have a height-time measurement from SOHO/LASCO at 17:06 UT: r = 23.56 Rs. Recall that the local sky-plane speed around this time is 2126 km s$^{-1}$, which can be used to extrapolate the distance to 17:00 UT as 22.46 Rs. Applying the projection correction, the LE height becomes r = 27.1 Rs, yielding L = 6.78 Rs. Using these f, df/dt, and L, we get the shock speed as V = 2643 km s$^{-1}$, in good agreement with CME LE speed (2564 km s$^{-1}$) from deprojected height-time measurements.

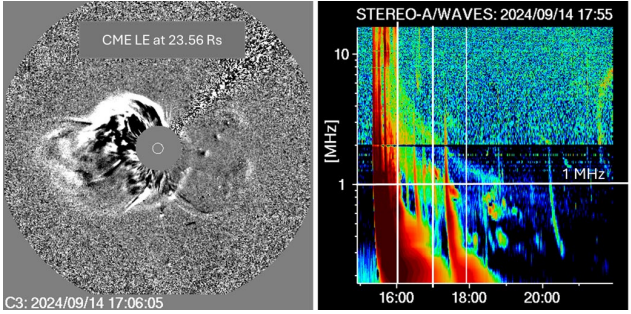

**Figure 5**. (left) SOHO/LASCO C3 difference image showing the CME with an LE height of 23.56 Rs at 17:06 UT in the LASCO C3 FOV. (right) STA/WAVES dynamic spectrum showing the type II burst with its harmonic structure. The type II burstwas faint but continued until about ~8:30 UT next day. The horizontal line at 1 MHz is drawn to indicate that the plasma frequency at 23.56 Rs from the Sun is ~1 MHz, corresponding to a local density of $1.2 \times 10^4$ cm$^{-3}$. The deprojected height should be 27.1 Rs for a source is at E56.

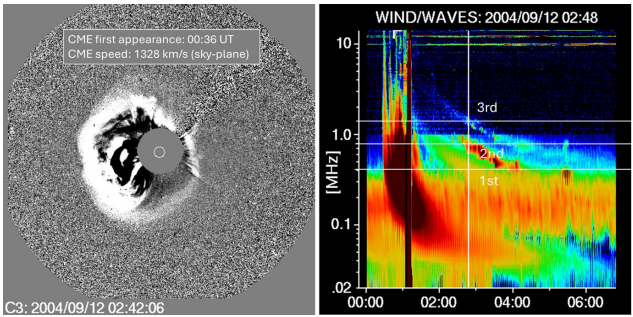

**Figure 5**. A fast (~1400 km s$^{-1}$) halo CME on 2004 September 12 (left) and the associated a broad-band (~0.6) type II burst observed by Wind/WAVES (right). The vertical white line marks the time of the SOHO/LASCO C3 frame on the left. The horizontal lines 0.4, 0.8, and 1.2 MHz show the first, second, and third harmonics. The 1$^{st}$ and 2nd harmonics merge with the background while the third harmonic is seen longer until ~400 kHz.

## 5. Discussion and Summary

One of the implications of the harmonic structure observed in the type II burst is the confirmation of the plasma emission mechanism. Bastian [14] suggested that type II-like bursts starting at frequencies >1 MHz and extending to frequencies below 1 MHz with smooth drift profiles in the dynamic spectrum are due to synchrotron emission by relativistic electrons interacting with high magnetic fields in under-dense plasma in the IP medium. The fundamental component of our type II burst is smooth, has large bandwidth, starts at frequencies > 1MHz, and extends to 0.17 MHz, except that it is accompanied by a second harmonic and possibly a third harmonic component. Figure 5 shows another IP type II satisfying all these criteria, but it has three clear harmonics as well. The harmonic structure cannot be explained by synchrotron mechanism. We do appreciate that the shock is extremely strong, accelerating protons to >300 MeV evidenced by the long-duration SGRE event and hence might accelerate electrons to very high energies as well. These electrons must be present upstream and downstream of the shock. The downstream is the sheath, which is denser than the upstream by a factor of 4. Since the synchrotron mechanism requires an under dense plasma, the sheath cannot be the region to produce synchrotron emission. The CME flux rope might have low density due to expansion, but the shock particles have no access to the flux rope. Flare electrons during the impulsive phase have access to the flux rope for a few minutes, but it is not clear if they survive for more than half a day in a rapidly expanding flux rope.

In summary, we reported on the third harmonic emission in IP type II bursts. The fundamental component starting below 10 MHz is smooth and of wide bandwidth; the burst extends to 0.17 MHz indicating a shock remaining strong far into the inner heliosphere. The shock speed derived from the drift rate of the type II burst under the plasma emission hypothesis is about the same as the CME speed derived from coronagraph observations. The existence of the harmonic structure is problematic for the synchrotron emission mechanism.

## 7. Acknowledgements

## References

[1] V. L. Bakunin et al. 1990, SolPhys 129, 379, doi: 10.1007/BF00159048.
[2] B. Kliem, A. Krueger, R. A. Treumann, 1992, SolPhys 140, 149, doi: 10.1007/BF00148435
[3] E. Ya. Zlotnik et al., 1998, A+A 331, 1087
[4] C. Kraft, P. Savioini, 2022, ApJL 934, L28, doi: 10.3847/2041-8213/ac7f28.
[5] G. A. Brueckner et al. 199y, SolPhys. 162, 3757, doi: 10.1007/BF00733434
[6] R. A. Howard et al., 2008, SSRv 136, 67, doi : 10.1007/s11214-008-9341-4
[7] J.-L. Bougeret et al., 1995, SSRv 71, 231. doi : 10.1007/BF00751331
[8] J.-L. Bougeret et al., 2008, SSRv 136, 487, doi : 10.1007/s11214-007-9298-8
[9] N. Gopalswamy, et al. 2012, SSRv 171, 23, doi : 10.1007/s11214-012-9890-4
[10] N. Gopalswamy, et al. 2021, ApJ 915, 82, doi : 10.3847/1538-4357/ac004f
[11] P. Makela et al. 2023, ApJ 954, 79, doi: 10.3847/1538-4357/ace627
[12] N. Gopalswamy, 2011, Planetary Radio Emission VII, p. 327
[13] N. Gopalswamy et al. 2003, GRL 30 (12), CiteID 8015, doi: 10.1029/2002GL016435
[14] T. Bastian, 2007, ApJ 665, 805, doi:10.1086/519246